\documentclass[pre,preprint,showpacs]{revtex4}
\usepackage{amsmath}
\usepackage{graphicx}

\begin{document}

\title{Abelian deterministic self organized criticality model: Complex dynamics
of avalanche waves. }

\author{Jozef \v{C}ern\'{a}k}

\email{jcernak@upjs.sk}

\affiliation{P. J. \v{S}af\'{a}rik University in Ko\v{s}ice, Institute of Physics,
Jesenn\'{a} 5, 04000 Ko\v{s}ice, Slovak Republic}


\begin{abstract}
The aim of this study is to investigate a wave dynamics and size scaling
of avalanches which were created by the mathematical model {[}J. \v{C}ern\'{a}k,
Phys. Rev. E \textbf{65}, 046141 (2002)]. Numerical simulations were
carried out on a two dimensional lattice $L\times L$ in which two
constant thresholds $E_{c}^{I}=4$ and $E_{c}^{II}>E_{c}^{I}$ were
randomly distributed. A density of sites $c$ with the threshold $E_{c}^{II}$
and threshold $E_{c}^{II}$ are parameters of the model. I have determined
autocorrelations of avalanche size waves, Hurst exponents, avalanche
structures and avalanche size moments for several densities $c$ and
thresholds $E_{c}^{II}$. I found correlated avalanche size waves
and multifractal scaling of avalanche sizes not only for specific
conditions, densities $c=0.0$, $1.0$ and thresholds $8\leq E_{c}^{II}\leq32$,
in which relaxation rules were precisely balanced, but also for more
general conditions, densities $0.0<c<1.0$ and thresholds $8\leq E_{c}^{II}\leq32$,
in which relaxation rules were unbalanced. The results suggest that
the hypothesis of a precise relaxation balance could be a specific
case of a more general rule. 
\end{abstract}
\maketitle

\section{Introduction\label{sec:Introduction}}

Bak, Tang, and Wiesenfeld (BTW) \cite{BTW} introduced a concept
of self-organized criticality (SOC) to study dynamical systems with
spatial degree of freedom. They proposed a simple model with conservative
and deterministic relaxation rules to demonstrate the SOC phenomenon.
Manna \cite{Manna} designed another conservative SOC model in which
stochastic relaxation rules were defined. Striving to find common
features of the models and to know their basic behaviors stimulated
many numerical and theoretical studies during the past two decades. 

Based on renormalization group calculations, Pietronero \textit{et
al}. \cite{Pietro} claimed that both deterministic \cite{BTW}
and stochastic models \cite{Manna} belong to the same universality
class, i.e. a small modification in the relaxation rules cannot change
universality class. Chessa \textit{et al}. \cite{Stanley} assumed
that finite size scaling (FSS) is common property of both deterministic
\cite{BTW} and stochastic \cite{Manna} models. With FSS the avalanche
size, area, lifetime, and perimeter follow power laws with cutoffs
\cite{Stanley}: 

\begin{equation}
P(x)=x^{-\tau_{x}}F(x/L^{D_{x}}),\label{eq:FSS}\end{equation}

where $P(x)$ is the probability density function of $x$, $F$ is
the cutoff function, and $\tau_{x}$ and $D_{x}$ are the scaling
exponents. The set of scaling exponents $(\tau_{x},\: D_{x})$ defines
the universality class \cite{Stanley}. A SOC model is \textit{Abelian}
if a final stable configuration (see below) does not depend on the
relaxation order. The BTW model is \textit{Abelian}, however the M
model is also \textit{Abelian} \cite{Dhar} only if we consider probabilities
of many stable configurations. 

Based on numerical simulations and an extended set of exponents, Ben-Hur
and Biham \cite{Hur} claimed that the BTW and M models cannot belong
to the same universality class. A precise numerical analysis of probability
density functions $P(x)$ led L\"{u}beck and Usadel \cite{Lubeck}
to the same conclusion. Essential progress to understand the discrepancy
between theoretical \cite{Pietro,Stanley} and numerical conclusions
\cite{Hur,Lubeck} has been achieved by Tebaldi \emph{et al.} \cite{Tebaldi}.
They found that avalanche size probability density functions $P(s)$
do not display FSS but show a multifractal scaling i.e. the avalanche
size exponent $\tau_{s}$ (Eq. \ref{eq:FSS}) does not apply to the
BTW model. Karmakar \emph{et al. }\cite{Karmakar_prl} proposed a
hypothesis that the presence or absence of a precise relaxation balance
between the amount released by a site and the total quantity which
the same site receives when all its neighbors relax at once determines
the appropriate universality class. Based on the precise relaxation
balance hypothesis \cite{Karmakar_prl} Karmakar and Manna \cite{Karmakar_pre}
proposed a flow chart to classify different SOC models into two universality
classes i.e. the BTW and Manna universality classes. 

The probability density functions of avalanche sizes $P(s)$ in Eq.
\ref{eq:FSS} show transitions from multifractal to FSS scaling for
certain densities of disturbing sites \cite{Karmakar_prl,Cernak_2006}.
The models \cite{Karmakar_prl,Cernak_2006} are stochastic and \textit{non-Abelian}
with unbalanced relaxation rules \cite{Karmakar_prl}. In this study,
I focus on verifying an existence of such transitions for the deterministic
and \textit{Abelian} model \cite{cer_2002} (In the orginal paper
\cite{cer_2002} the model was incorrectly clasified as non-\textit{Abelian})
with unbalanced relaxation rules. The model \cite{cer_2002} displays
an anomalous increase of the avalanche size area exponents $\tau_{a}$
(Eq. \ref{eq:FSS}) for densities near $c=0.01$ and thresholds $E_{c}^{II}\geq16$.
However, the cause of this anomalous behavior is not well understood.
I assumed that the transition from multifractal to FSS scaling could
take place for density $c<0.01$ and threshold $E_{c}^{II}\geq16$,
because relaxation rules change character from balanced ($c=0.0$)
to unbalanced ($c>0.0$). To characterize avalanche size scaling I
investigated avalanche wave dynamics \cite{Menech,Stella}, Hurst
exponents \cite{Menech}, avalanche structures \cite{Hur} and avalanche
size moments \cite{Menech}. 

The paper is organized as follows. In Sec. \ref{sec:model} I repeat
a definition of the inhomogeneous sand pile model \cite{cer_2002}.
In Sec. \ref{sec:Results} I determine autocorrelations and fluctuations
of avalanche size waves, avalanche structures and avalanche size moments
for several densities $0.0\leq c$$\leq1.0$ and thresholds $8\leq E_{c}^{II}\leq32$.
Sec. \ref{sec:Discussion} is devoted to a discussion which is followed
by conclusions in Sec. \ref{sec:Conclusion}.

\section{\label{sec:model} An Abelian deterministic and conservative self
organized criticality model}

The inhomogeneous sand pile model \cite{cer_2002} is defined on
a two dimensional lattice of size $L\times L$. Each site $\mathbf{i}$
has assigned variables $E(\mathbf{i})$ and $E_{c}(\mathbf{i})$.
The variable $E(\mathbf{i})$ is dynamical and it represents a physical
quantity such as energy, grain density, and etc. The threshold $E_{c}(\mathbf{i})$
is a static value at site $\mathbf{i}$ which is defined only once
during initialization of simulations. The threshold $E_{c}(\mathbf{i})$
has two values \cite{cer_2002}:

\begin{eqnarray}
E_{c}(\mathbf{i}) & = & \begin{cases}
E_{c}^{I}=2d\\
E_{c}^{II}=2dk, & k=2,3,4,\ldots,\end{cases}\end{eqnarray}

where $d$ is a dimension and $k$ is a natural number. The model
has two parameters namely the density $c=n/L^{2}$ and threshold $E_{c}^{II}$
where $n$ is a number of sites with the threshold $E_{c}(\mathbf{i})=E_{c}^{II}$,
remaining $L^{2}-n$ sites have the threshold $E_{c}^{I}=4$. During
initializations of simulations, $n$ sites with thresholds $E_{c}(\mathbf{i})=E_{c}^{II}$
were picked out randomly and all remaining sites had the threshold
$E_{c}^{II}=4$, thus a set of the thresholds $\left\{ E_{c}(\mathbf{i})\right\} $
represents a quenched disorder. A stable configuration is defined
by a condition $E(\mathbf{i})<E_{c}(\mathbf{i})$ for each site $\mathbf{i}$.
Let us assume that from a stable configuration we iteratively select
$\mathbf{i}$ at random and increase $E(\mathbf{i})$$\rightarrow E(\mathbf{i})+1$.
If an unstable configuration is reached i.e. $E(\mathbf{i})\geq E_{c}(\mathbf{i})$
then a relaxation starts. The relaxation rules are conservative and
deterministic \cite{cer_2002}:

\begin{equation}
E(\mathbf{i})\rightarrow E(\mathbf{i})-\sum_{e}\triangle E(\mathbf{e}),\label{eq:r1}\end{equation}

\begin{equation}
E(\mathbf{i}+\mathbf{e})\rightarrow E(\mathbf{i}+\mathbf{e})+\triangle E(\mathbf{e}),\label{eq:r2}\end{equation}

\begin{equation}
\sum_{e}\triangle E(\mathbf{e})=E_{c}(\mathbf{i}),\label{eq:r3}\end{equation}

where $\mathbf{e}$ is a set of vectors from the site $\mathbf{i}$
to its neighbors. The relaxation rules (Eqs. \ref{eq:r1}-\ref{eq:r3})
are repeated until the site $\mathbf{i}$ becomes stable. If the neighbors
of the site $\mathbf{i}$ become unstable then avalanche can run on.
All unstable sites belong to the avalanche. The relaxations given
by Eqs. \ref{eq:r1}-\ref{eq:r3} are repeated until a stable configuration
is reached, i.e. $E(\mathbf{i})<E_{c}(\mathbf{i})$ for all sites
$\mathbf{i}$. Stable and unstable configurations are repeated many
times. A total number of relaxations during one avalanche is an avalanche
size $s.$

\section{\label{sec:Results}Numerical results}

Numerical simulations were carried out on two dimensional lattices
$L\times L$ where the linear lattice size $L$ was $L=128,\:256,\:512,\:1024,\:2048,$
and $4096$. A density $c$ and thresholds $E_{c}^{I}=4$, $E_{c}^{II}=8,\:16$
and $32$ were chosen based on the previous results \cite{cer_2002}
to cover a parameter space in which interesting behaviors were expected
(Sec. \ref{sec:Introduction}). For example, near the density $c=0.01$
an anomalous increase of the avalanche area scaling exponent $\tau_{a}$
(Eq. \ref{eq:FSS}) has been observed \cite{cer_2002}. Near densities
$c=0.0$ and $1.0$ local relaxation rules change character from balanced
($c=0.0$ and $1.0$) to unbalanced ($c>0.0$ and $c<1.0$) \cite{Karmakar_prl,Karmakar_pre},
thus a transition from multifractal to finite size scaling could take
place in the intervals $c>0.0$ and $c<1.0$. Considering these assumptions,
I have selected the density $c$ as follows: $c=a$ (surroundings
of $c=0.0$) or $c=1.0-a$ (surroundings of $c=1.0$) where $a=0,\:0.001,\:0.002,\:0.004,\;0.008,\;0.01,\:0.02,\:0.04,$
and $0.08$. In addition, to cover the whole interval $0.0<c<1.0$,
I added the sample concentrations $c=l/10$, where $l=1,\:2,\ldots,\;9$.
I have recorded about $10^{6}$ avalanches after initializations of
simulations in which an avalanche dynamics has to reach the SOC state
\cite{BTW}. To qualify a reproducibility of the results all numerical
simulations were repeated once for each lattice size $L\times L$,
concentration $0.0\leq c$$\leq1.0$ and threshold $E_{c}^{II}$.
A comparison of these data sets showed that the results are well reproducible.

\begin{figure*}
\includegraphics[width=6cm]{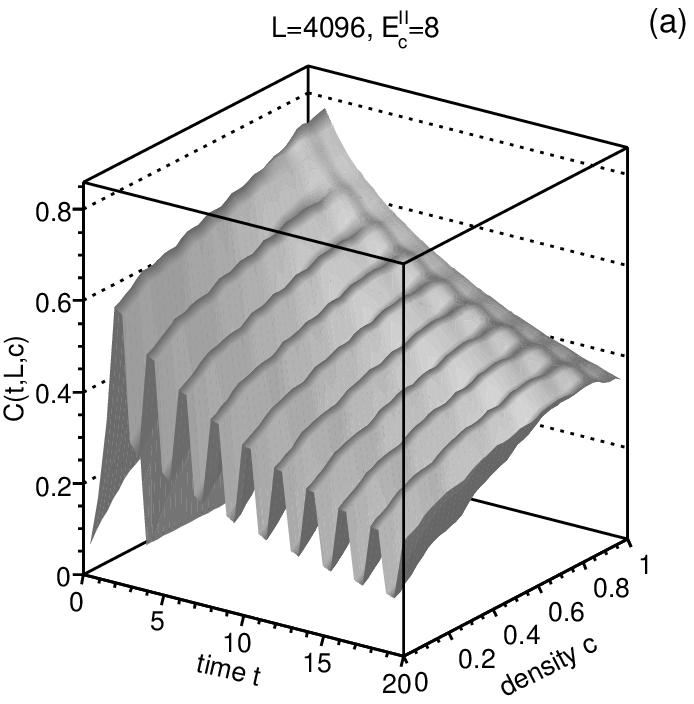}\includegraphics[width=6cm]{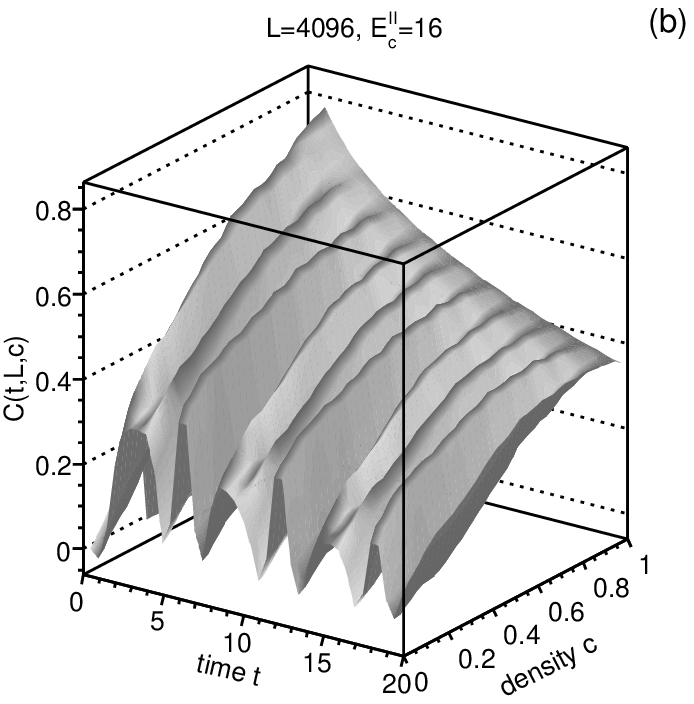}\includegraphics[width=6cm]{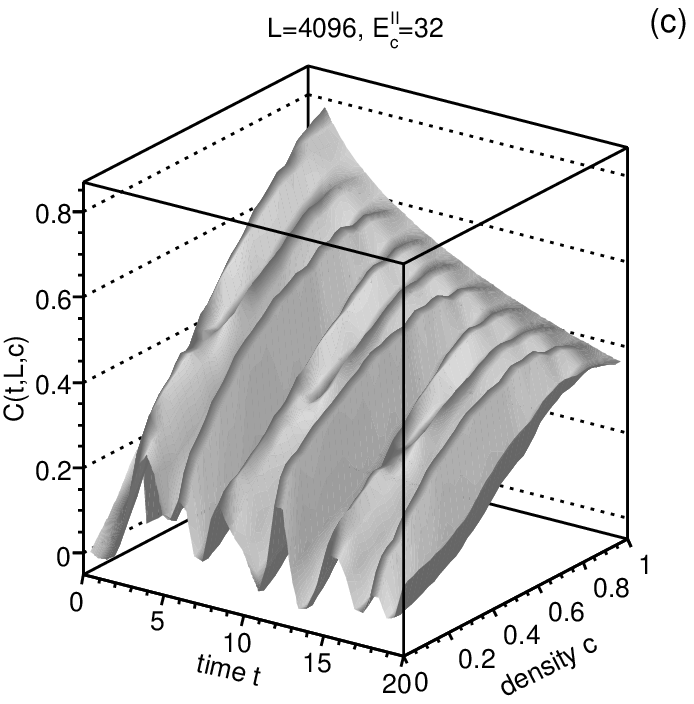}

\caption{\label{fig:Autocorrelation}Autocorrelations $C(t,\: L,\: c)$ for
the linear lattice size $L=4096$, for densities $0.001\leq c\leq1.0$
and for thresholds $E_{c}^{I}=4$, (a) $E_{c}^{II}=8$, (b) $E_{c}^{II}=16$,
and (c) $E_{c}^{II}=32$. The autocorrelations $C(t,L,c=0)$ are not
shown.}

\end{figure*}

A possibility to decompose an avalanche into waves \cite{Ivash}
is a significant advantage of computer models. Because avalanche wave
dynamics \cite{Menech,Stella} can provide valuable initial information
about the character of an SOC model. An avalanche of size $s$ is
decomposed into $m$ waves with size $s_{k}$, where $s=\sum_{k=1}^{m}s_{k}$.
A time sequence of avalanche waves $s_{k}$ is used to determine the
autocorrelation function \cite{Menech,Stella}

\begin{equation}
C(t,\: L)=\frac{\langle s_{k+t}s_{k}\rangle_{L}-\langle s_{k}\rangle_{L}^{2}}{\langle s_{k}^{2}\rangle_{L}-\langle s_{k}\rangle_{L}^{2}},\end{equation}

where time is $t=1,\:2,\ldots,$ and the time averages are taken over
$5\times10^{6}$ waves. Autocorrelations $C(t,\: L,\: c)$ have been
analyzed for the time $1\leq t\leq1000$, lattice sizes $128\leq L\leq4096$,
selected concentrations $0.0\leq c\leq1.0$ and thresholds $E_{c}^{I}=4,$
$E_{c}^{II}=8,\:16$, and $32$ (see above). The autocorrelations
$C(t,\: L=4096,\: c)$ for the biggest lattice size $L=4096$ are
shown in Fig. \ref{fig:Autocorrelation}. I have observed that for
the density $c=1.0$, the autocorrelations $C(t,\: L=4096,\: c=1.0)$
agree within experimental error with autocorrelations of the BTW model
\cite{Menech} ($C(t,\: L=4096,\: c=0.0)$) . I note that the autocorrelations
$C(t,\: L=4096,\: c=0.0)$ are not shown in Fig. \ref{fig:Autocorrelation}.
I have approximated the autocorrelations $C(t,\: L,\: c=0,\:1.0)$
by a simple function $C(t,L,c=0,\:1.0)\sim\exp(-\alpha t)$ where
$\alpha$ is a decay rate \cite{cer_2008}. For more general conditions,
densities $0.0<c<1.0$ (Fig. \ref{fig:Autocorrelation}), the autocorrelations
$C(t,\: L,\: c)$ are more complex functions than for specific densities
$c=0.0$ and $1.0$. I have found that with increasing time $t$,
the autocorrelations $C(t,\: L,\: c)$ are decreasing functions. An
unexpected finding is the existence of oscillating components of autocorrelations.
For all densities $0.0<c<1.0$ (Fig. \ref{fig:Autocorrelation}(a))
and the threshold $E_{c}^{II}=8$ their periods are approximately
constant. Amplitudes of oscillating components decrease with increasing
density $c$ and time $t$. At the given time $t$ the autocorrelations
$C(t,\: L,\: c)$ increase if a density $c$ increases, i.e. if $c_{2}\geq c_{1}$
then $C(t,\: L,\: c_{1})\geq C(t,\: L,\: c_{2})$. Near densities
$c=0.0$ and $1.0$, the oscillating parts of $C(t,\: L,\: c)$ disappear.
I have observed more complex behaviors (Fig. \ref{fig:Autocorrelation}
(b) and (c)) for thresholds $E_{c}^{II}\geq16$ than for the threshold
$E_{c}^{II}=8$ ((Fig. \ref{fig:Autocorrelation} (a)). The oscillating
components of $C(t,\: L,\: c)$ (Fig. \ref{fig:Autocorrelation}(b))
have longer periods for densities $c<0.08$ and threshold $E_{c}^{II}=16$
than for the threshold $E_{c}^{II}=8$. However, odd periods were
split for densities $c>0.08$. The same effect take place for the
threshold $E_{c}^{II}=32$, but a critical density is higher $c\doteq0.40$
(Fig. \ref{fig:Autocorrelation}(c)). After splitting the oscillating
components, for thresholds $E_{c}^{II}=16$ and $32$, their new periods
were approximately equal to the period which has been found for the
threshold $E_{c}^{II}=8$.

Stochastic processes are often characterized by Hurst exponents \cite{Mandelbrot}.
To determine the Hurst exponent the fluctuation $F(t,\: L)$ \cite{Menech}: 

\begin{equation}
F(t,\: L)=\left[\langle\Delta y(t)^{2}\rangle_{L}-\langle\Delta y(t)\rangle_{L}^{2}\right]^{1/2}\end{equation}

is used where $y(t)=\sum_{k=1}^{t}s_{k}$ and $\Delta y(t)=y(k+t)-y(k)$.
If a fluctuation $F(t,\: L)$ scales with the time $t$ as $F(t,\: L\rightarrow\infty)\sim t^{H}$
then $H$ is the Hurst exponent.

\begin{figure}
\includegraphics[width=7cm]{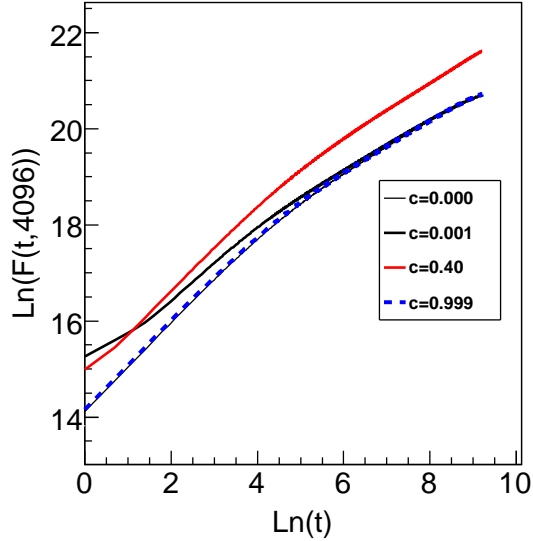}

\caption{\label{fig:Fluct.}(Color online) Fluctuations $F(t,\: L=4096)$ for
densities $c=0.000,\:0.001,\:040$ and $0.999$ and the threshold
$E_{c}^{II}=32$.}

\end{figure}

\begin{figure*}
\includegraphics[width=6cm]{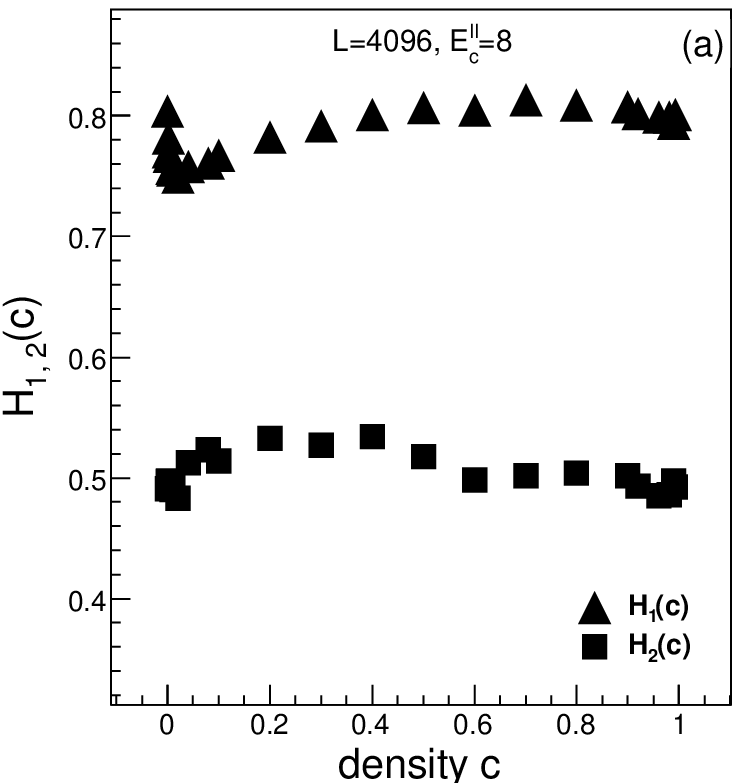}\includegraphics[width=6cm]{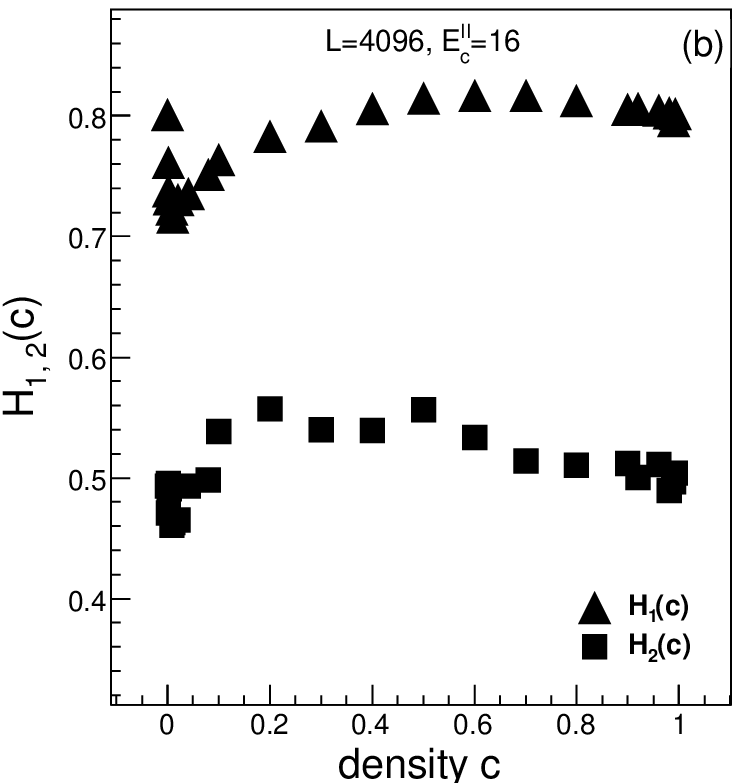}\includegraphics[width=6cm]{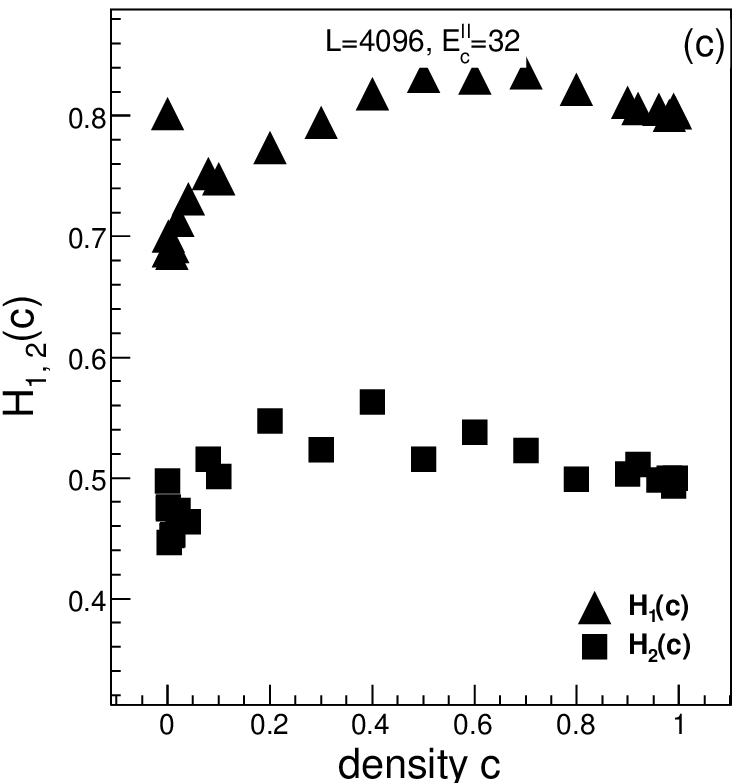}

\caption{\label{fig:Hurst}Hurst exponents $H_{1}(c)$and $H_{2}(c)$ as functions
of a densities $c$ and thresholds: $E_{c}^{I}=4$, (a) $E_{c}^{II}=8$,
(b) $E_{c}^{II}=16$, and (c) $E_{c}^{II}=32$ for the lattice size
$4096\times4096$. The Hurst exponents at given density $c$, $H_{1}(c)$
and $H_{2}(c)$ were approximated using the power laws $F(t,\: L)\sim t^{H_{1}(c)}$
and $F(t,\: L)\sim t^{H_{2}(c)}$ in the intervals $t=1-100$ and
$t=10^{3}-10^{4}$. }

\end{figure*}

I have determined fluctuations and the corresponding Hurst exponents
for the lattice size $L=4096$, selected concentrations $c$ (see
above) and thresholds $E_{c}^{I}=4,$ $E_{c}^{II}=8,\:16$, and $32$.
For all these parameters fluctuations show two scaling regions $1<t_{1}<100$
and $1000<t_{2}<10000$. The fluctuations $F(t,\: L)$, for densities
$c=0.000,\:0.001,\:040$ and $0.999$ and threshold $E_{c}^{II}=32$,
are shown in Fig. \ref{fig:Fluct.} to demonstrate the existence of
two scaling intervals. The Hurst exponents $H_{1}(c)$ and $H_{2}(c)$
as functions of density $c$ and threshold $E_{C}^{II}$ are shown
for all parameters (densities $0.0\leq c\leq1.0$ and thresholds $8\leq E_{c}^{II}\leq32$)
in Fig. \ref{fig:Hurst}. I have observed that the exponents $H_{1}(c)$
and $H_{2}(c)$ depend on the parameters (density $c$ and threshold
$E_{C}^{II}$) in a nontrivial manner (Fig. \ref{fig:Hurst}). For
densities $0.0\leq c\leq1.0$ and thresholds $8\leq E\leq32$ functions
$H_{1}(c)$ are bounded by the interval $0.68<H_{1}(c)<0.81$. Similarly,
the functions $H_{2}(c)$ are limited by the interval $0.44<H_{2}(c)\leq0.56$.
The exponents $H_{1}(c)\doteq0.80$ and $H_{2}(c)\doteq0.50$ are
approximately constant for densities $c>0.50$. I have observed anomalous
decreases of functions $H_{1,\,2}(c)$ near low densities $0.0<c\leq0.01$
(Fig. \ref{fig:Hurst}). In addition, functions of $H_{1\,,2}(c)$
have a decreasing tendency if the second thresholds $E_{c}^{II}$
increase. Finally, the functions $H_{1,2}(c)$ are not symmetric around
the density $c=1/2$, i.e. $H_{1}(0.5-a)\neq H_{1}(0.5+a)$ and $H_{2}(0.5-a)\neq H_{2}(0.5+a)$
for $0.0<a<0.5$ except the specific densities $c=0.0$ and $c=1.0$,
where $H_{1}(0.0)\doteq H_{1}(1.0)$ and $H_{2}(0.0)\doteq H_{2}(1.0)$
within experimental errors.

\begin{figure*}
\includegraphics[width=4cm,angle=90]{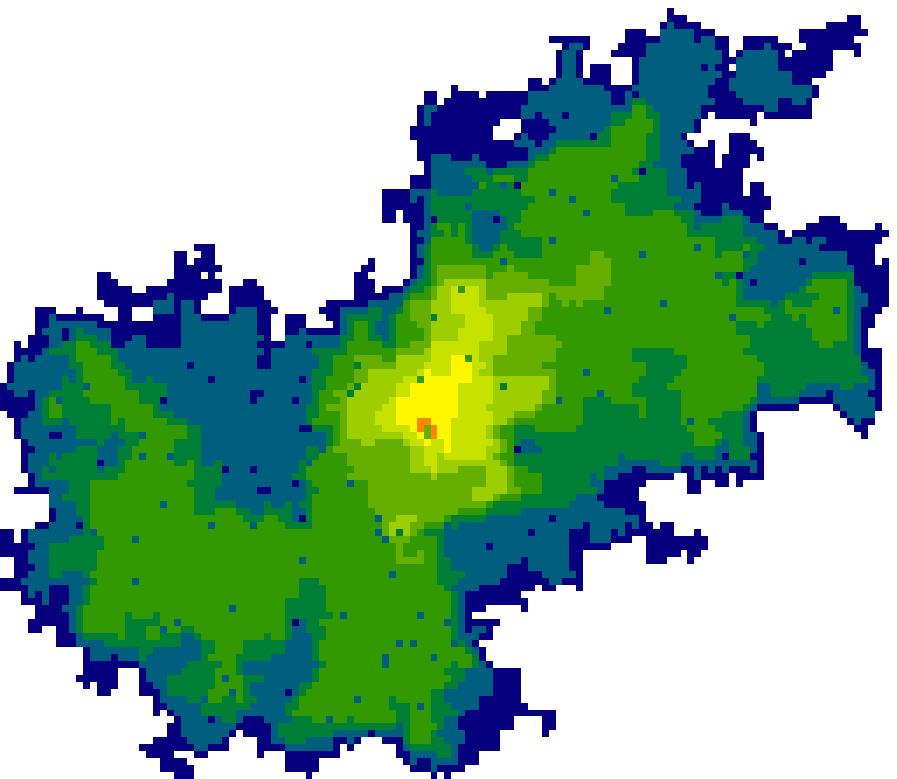}(a) \includegraphics[width=4cm]{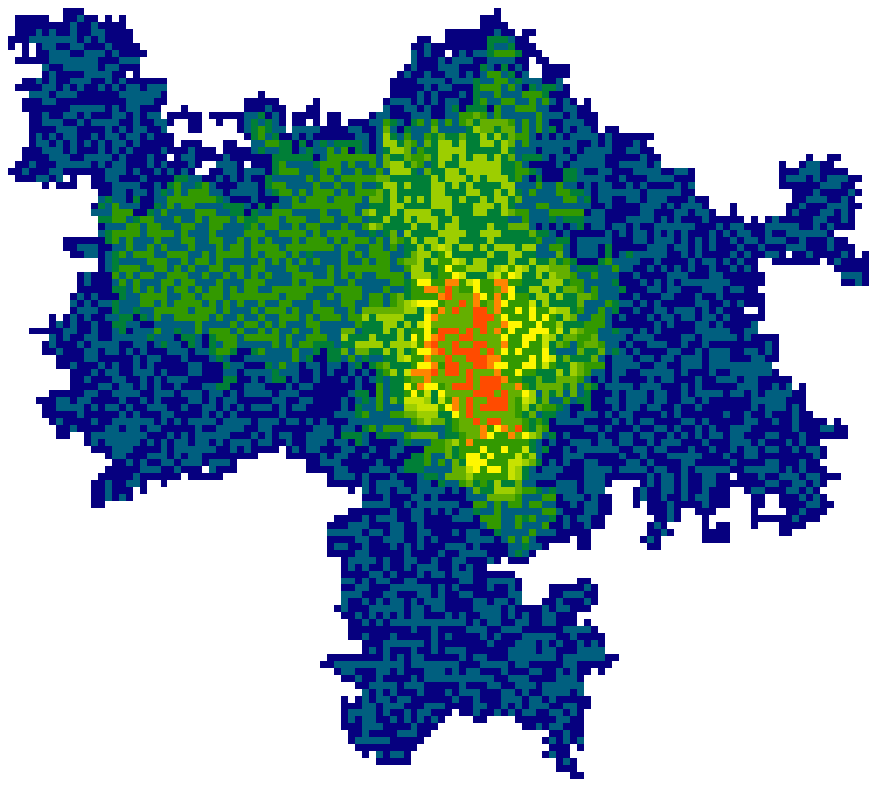}(b)
\includegraphics[width=4cm]{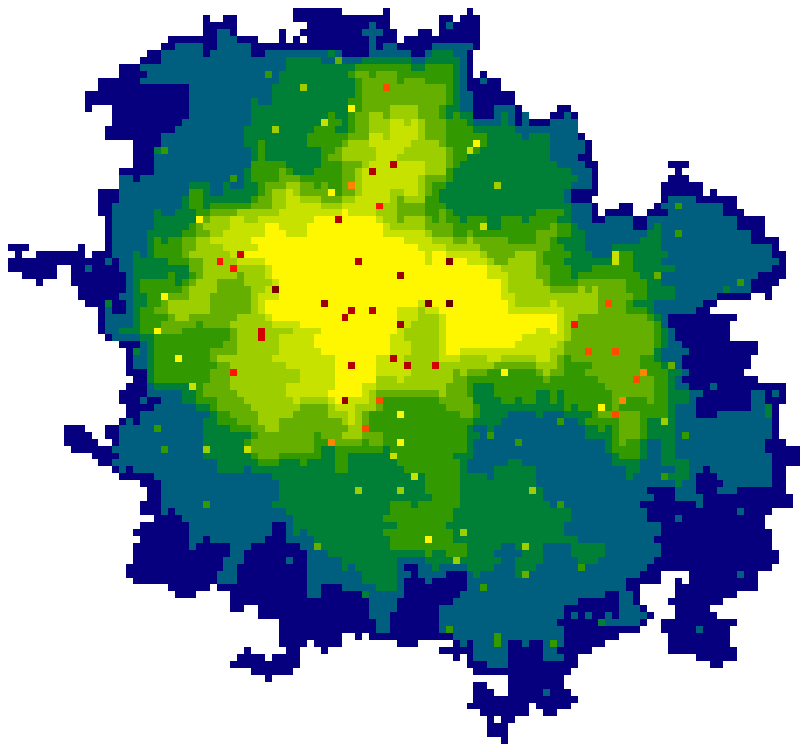}(c)

\includegraphics[width=4cm]{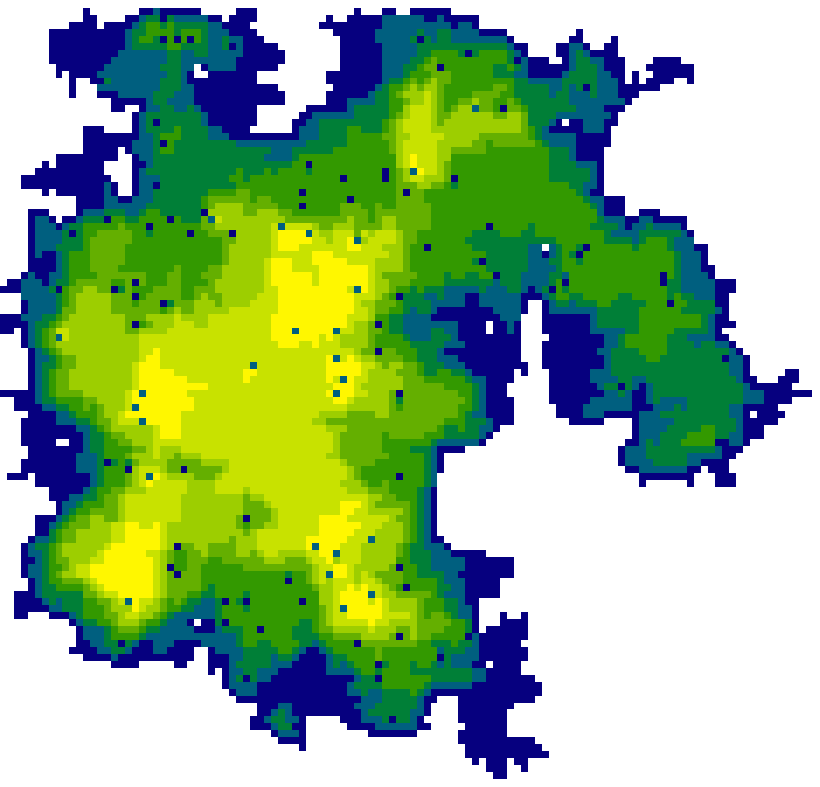}(d) \includegraphics[width=4cm]{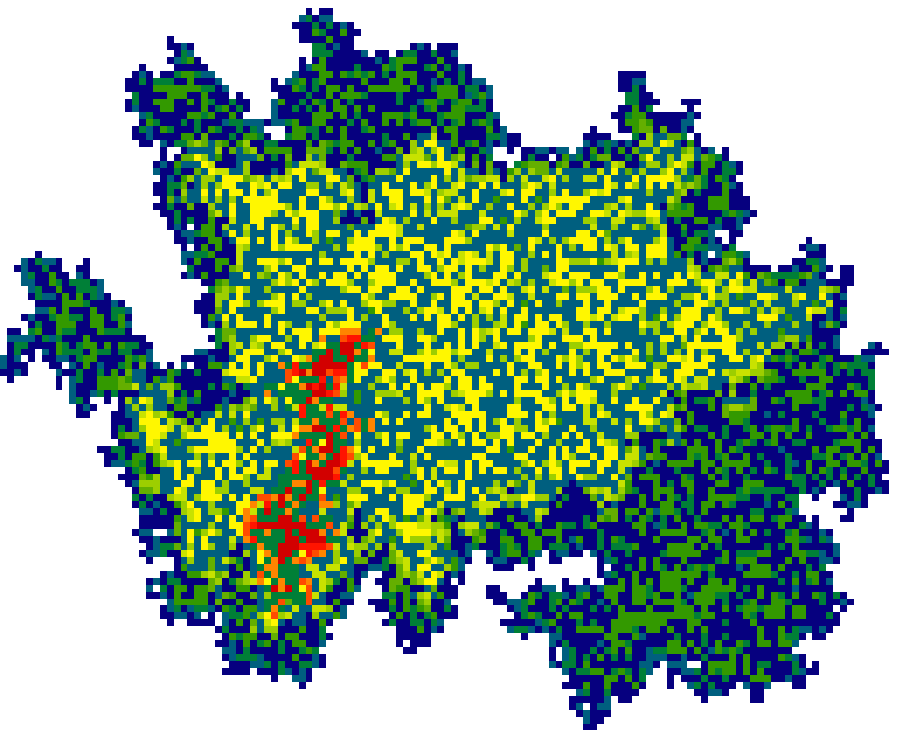}(e)
\includegraphics[width=4cm]{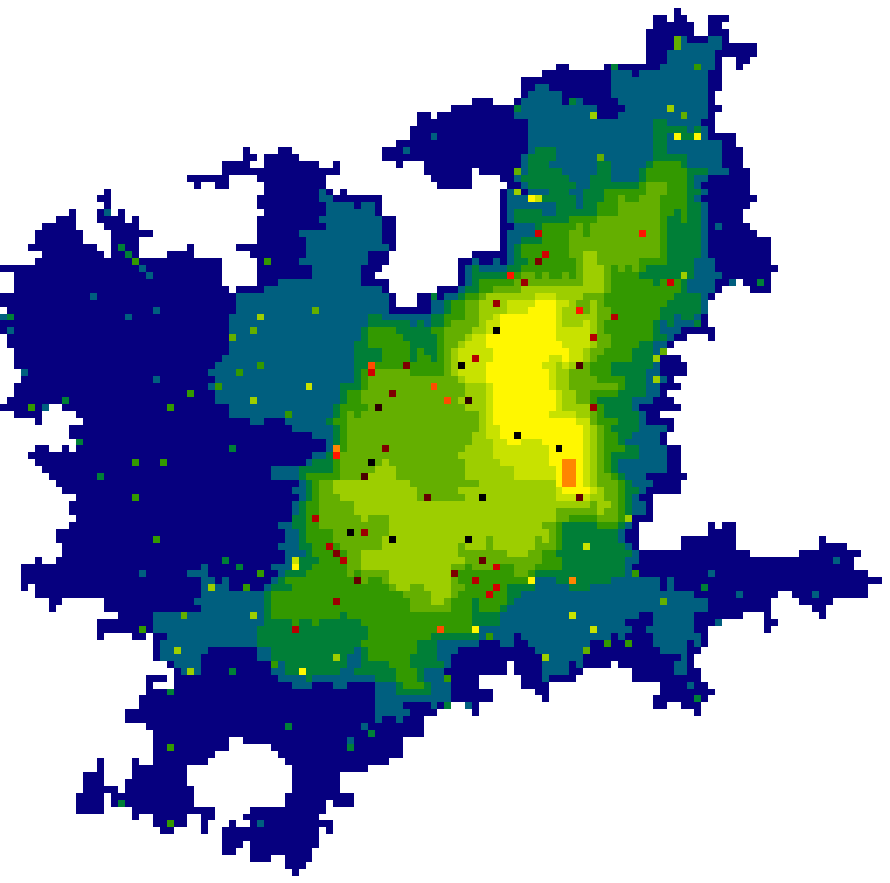}(f)

\includegraphics[width=4cm]{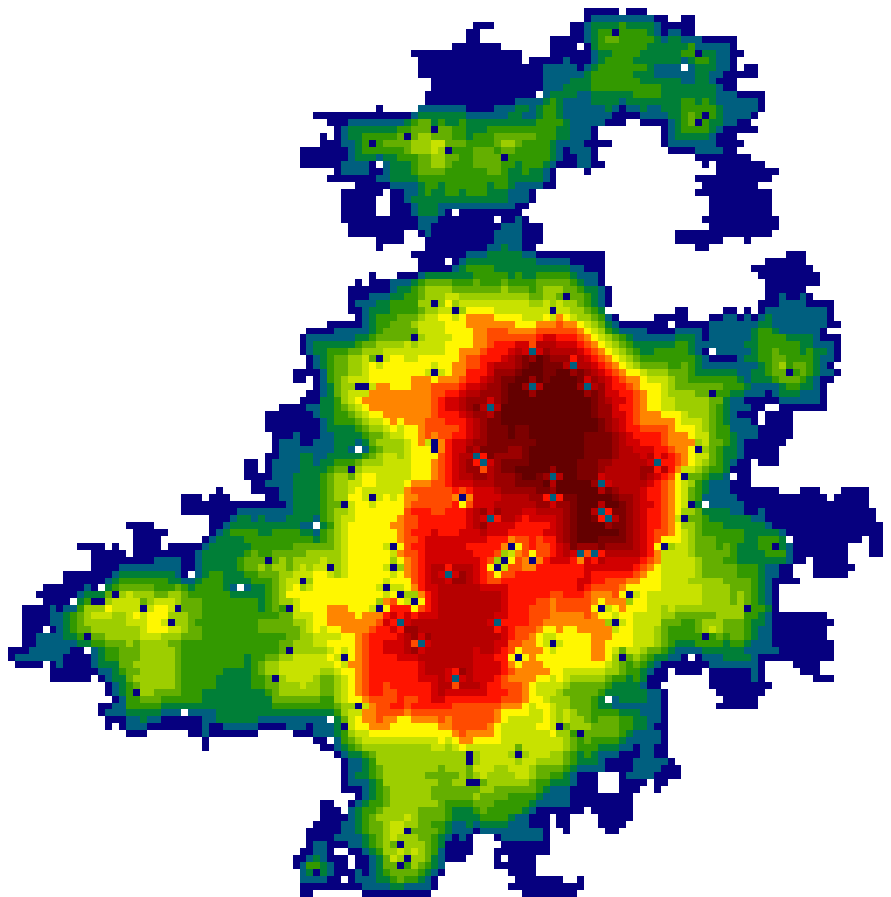}(g) \includegraphics[width=4cm]{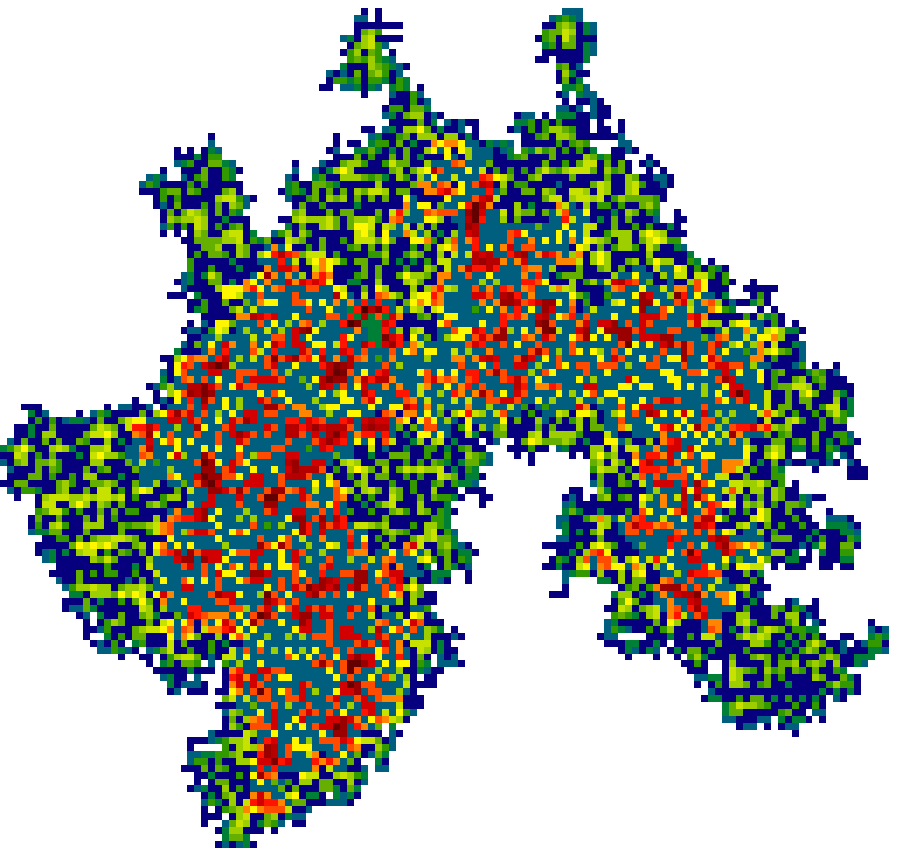}(h)
\includegraphics[width=4cm]{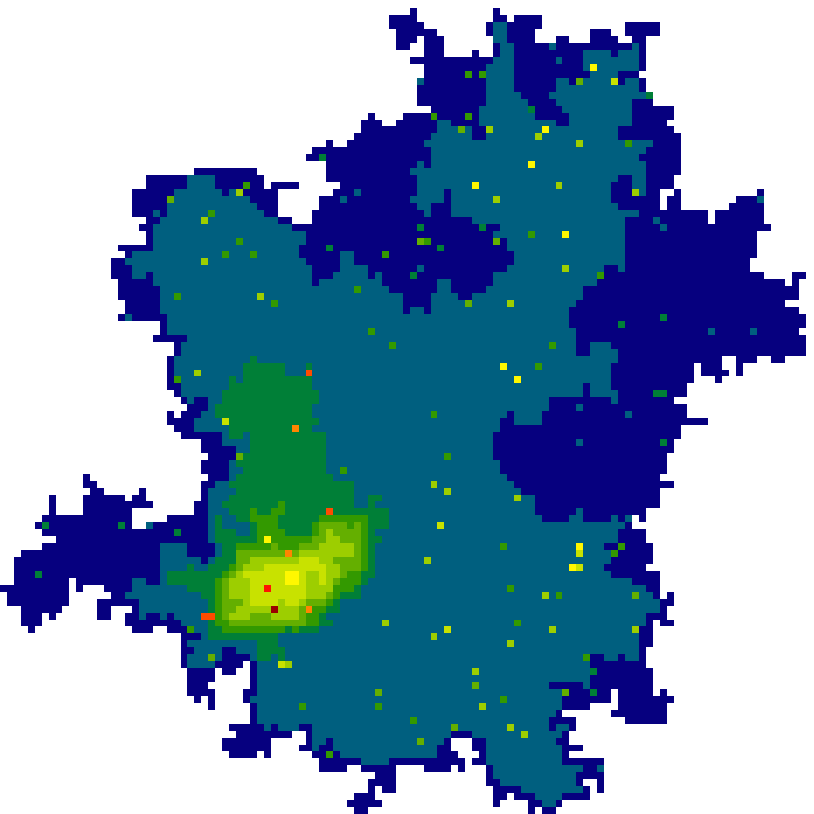}(i)

\caption{\label{fig:avalanche}Avalanche structures on two dimensional lattice
$128\times128$ for parameters $E_{c}^{I}=4$, (a) $E_{C}^{II}=8$,
$c=0.02$, (b) $E_{C}^{II}=8$, $c=0.5$, (c) $E_{C}^{II}=8$, $c=0.98$,
(d) $E_{C}^{II}=16$, $c=0.02$, (e) $E_{C}^{II}=16$, $c=0.5$, (f)
$E_{C}^{II}=16$, $c=0.98$, (g) $E_{C}^{II}=32$, $c=0.02$, (h)
$E_{C}^{II}=32$, $c=0.5$, and (i) $E_{C}^{II}=32$, $c=0.98$. Lattice
sites with the same numbers of relaxations are shown by the same color
(rainbow pseudo-color coding).}

\end{figure*}

Ben-Hur and Biham \cite{Hur} proposed to use avalanche structures
to demonstrate a difference between BTW \cite{BTW} and M \cite{Manna}
models. An avalanche structure consists of clusters of sites with
equal numbers of relaxations. The BTW model displays rigorous shell-like
structures \cite{Hur,Karmakar_prl} and the M model displays disordered
structures \cite{Hur} with inner holes \cite{Karmakar_prl}. I
have analyzed several avalanche structures (Fig. \ref{fig:avalanche})
of the inhomogeneous sand pile model \cite{cer_2002} to compare
them with known structures \cite{Hur,Karmakar_prl}. 

I have observed the avalanche structures which resemble the shell-like
structure for densities $c=0.02$ (Fig. \ref{fig:avalanche}(a), (d),
and (g)) and $c=0.98$ (Fig. \ref{fig:avalanche} (c), (f), and (i)).
However, these structures are not exactly shell-like. A clear visible
difference is the existence of holes in an avalanche, for example
see Figs. \ref{fig:avalanche}(a), (b), and (d). The sizes of these
holes vary from one site (obviously a site with $E_{c}^{II}>E_{c}^{I}$)
to several sites. Mainly, for low densities ($c=0.02)$ an existence
of holes is clearly demonstrated in Figs. \ref{fig:avalanche}(a),
(d), and (g) where the sites with the threshold $E_{c}^{II}$ can
absorb and relax more energy than surrounding sites with $E_{c}^{I}$.
If the sites ($E_{c}^{II}$) absorb energy then they are well identified
as small holes inside the avalanche structure (Figs. \ref{fig:avalanche}(a),
(d), and (g)). If the sites ($E_{c}^{II}$) release more energy than
their neighbors (with threshold $E_{c}^{I}$) then the sites ($E_{c}^{II}$)
involve instabilities of many sites within a certain distance. These
sites must relax to be stable (Fig. \ref{fig:avalanche} (d)). At
high density $c=0.98$, the sites with the lower threshold $E_{c}^{I}$
are considered for disturbing sites. A site with threshold $E_{c}^{I}$
can receive more energy from neighbors than a critical amount of the
site (threshold $E_{c}^{I}$). Then the site ($E_{c}^{I}$) must relax
more times than neighbors (sites with the threshold $E_{c}^{II}$)
to be stable. The disturbing sites are shown as isolated sites in
Figs. \ref{fig:avalanche}(c), (f), and (i). The effects of disturbing
sites $E_{c}^{II}$ and $E_{c}^{I}$ differs, for the small density
$c=0.02$, the disturbing sites $E_{c}^{II}$ can absorb and relax
more energy as their neighbors. However, for high density $c=0.98$,
the disturbing sites $E_{c}^{I}$ can only do more relaxations than
their neighbors sites. 

I have found new avalanche structures which resemble neither shell-like
(the BTW mode) nor disordered (the M model) \cite{Hur,Karmakar_prl}
for density $c=0.50$ and thresholds $8\leq E_{c}^{II}\leq32$ Figs.
\ref{fig:avalanche}(b), (e), and (h). Their typical feature is the
existence of complex clusters in avalanche which more resemble percolation
clusters. 

The model (Sec. \ref{sec:model}) displays shell-like avalanche structures
as well as the BTW model \cite{Hur,Karmakar_prl} only for the specific
densities $c=0.0$ and $c=1.0$. 

\begin{figure*}
\includegraphics[width=6cm]{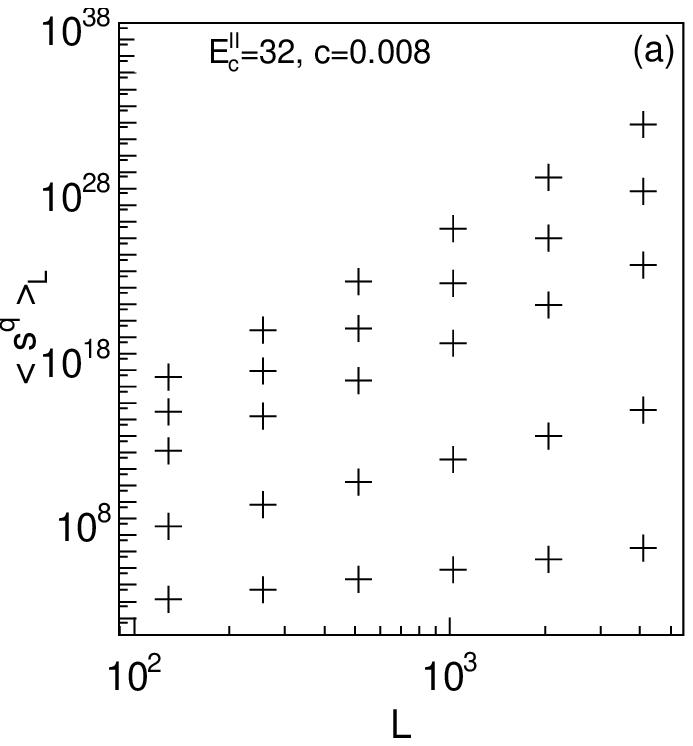}\includegraphics[width=6cm]{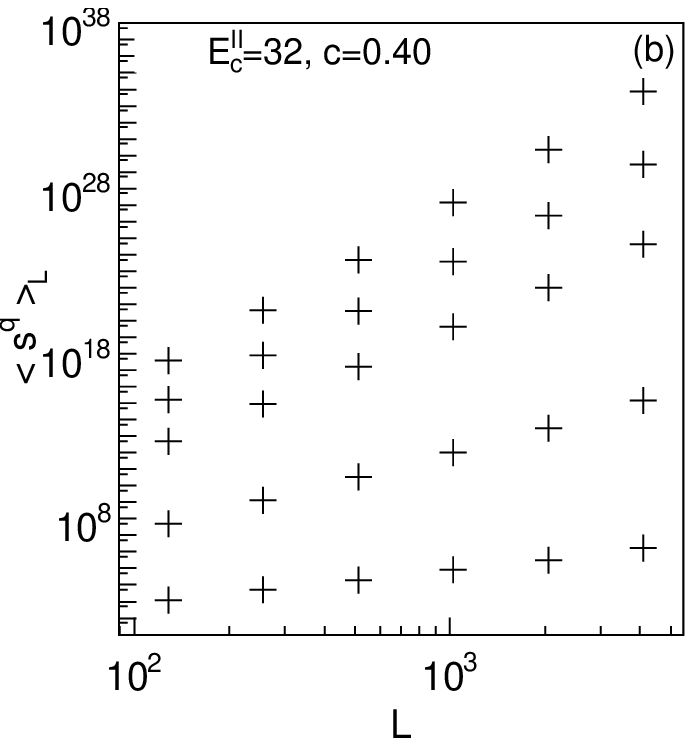}\includegraphics[width=6cm]{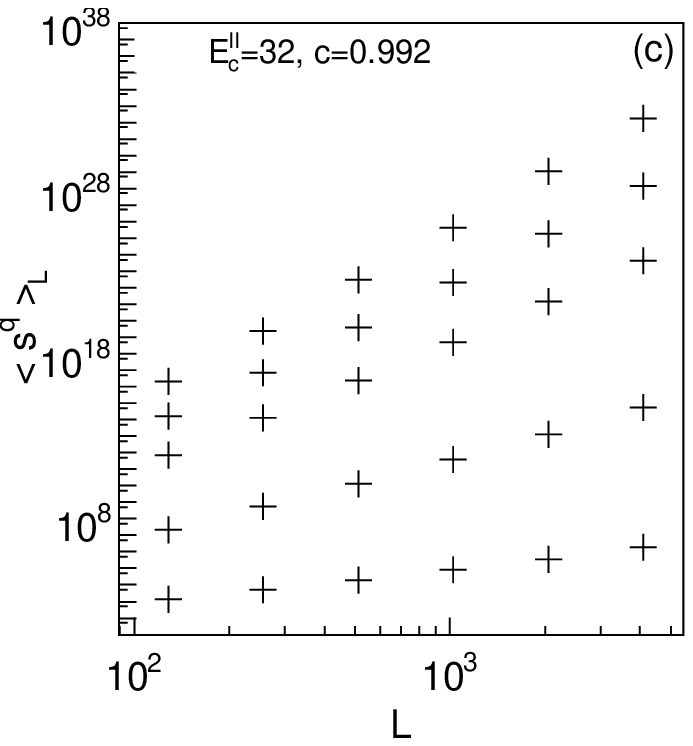}

\caption{\label{fig:scal}Scaling of avalanche size moments $\langle s^{q}\rangle$
versus the lattice size $128\leq L\leq4096$. The exponents $q$ are
$q=1.0,\:2.0,\:3.0,\:3.5$ and $3.95$ from bottom to top. The model
parameters are $E_{c}^{I}=4$, $E_{c}^{II}=32$, (a) $c=0.008$, (b)
$c=0.40$ and (c) $c=0.992$.}

\end{figure*}

Sometimes we cannot decompose an avalanche into waves, obviously if
we study real systems, then avalanche moments \cite{Menech} are
useful. A property $x$ in FSS obeys scaling given by Eq. \ref{eq:FSS}.
The moments $q$ are \cite{Menech}:

\begin{equation}
\langle x^{q}\rangle=\int_{_{0}}^{x_{max}}x^{q}P(x,L)dx\sim L^{\sigma_{x}(q)}\label{eq:moment}\end{equation}
 where $\sigma_{x}(q)=(q+1-\tau_{x})D_{x}$ and $x_{max}\sim L^{D_{x}}$.
I have determined only avalanche size moments $\langle s^{q}\rangle$
versus the lattice size $L$ which are shown in Fig. \ref{fig:scal}
for densities $c=0.008$, $c=0.40$ and $c=0.992$ and for thresholds
$E_{c}^{I}=4$ and $E_{c}^{II}=32$. The moments $\langle s^{q}\rangle$
scale with the lattice size $L$ as well as $\langle s^{q}\rangle\sim L^{\sigma_{x}(q)}$
(Eq. \ref{eq:moment}), thus a basic requirement is met to determine
$\sigma_{s}(q)$. 

\begin{figure*}
\includegraphics[width=6cm]{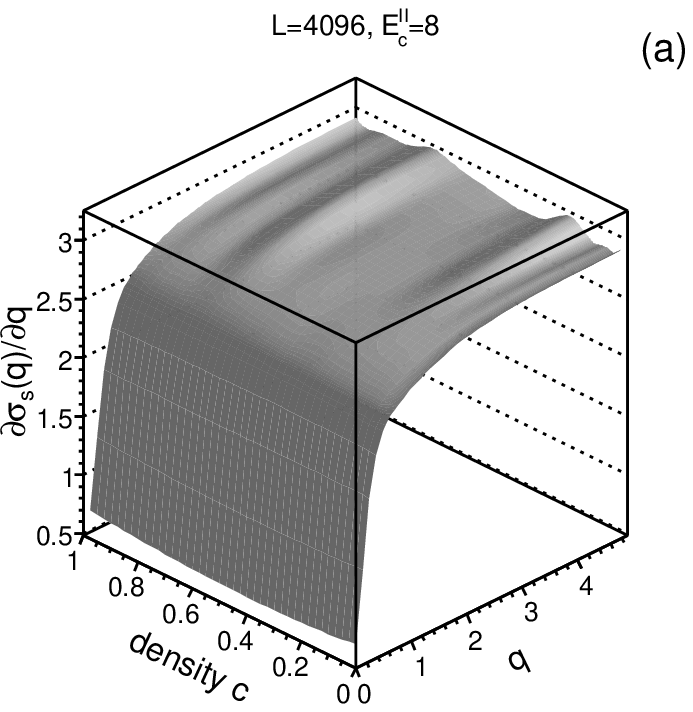}\includegraphics[width=6cm]{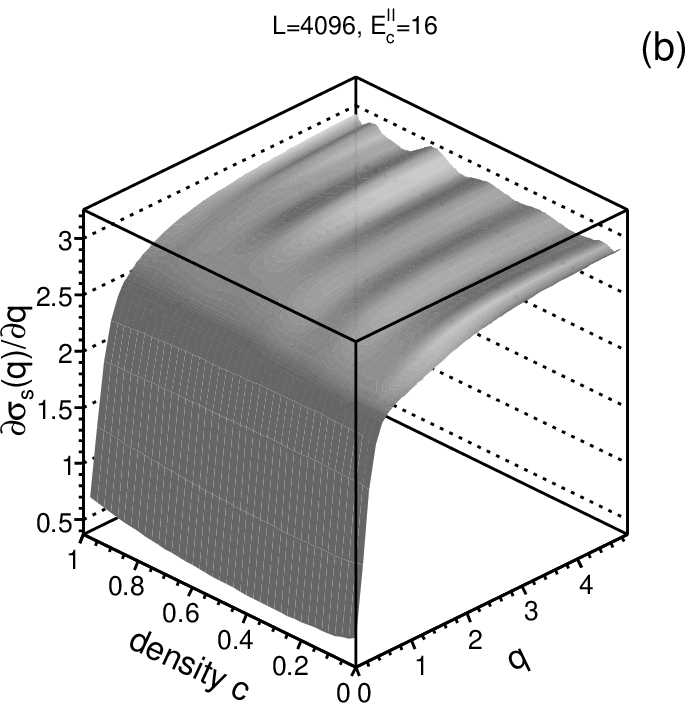}\includegraphics[width=6cm]{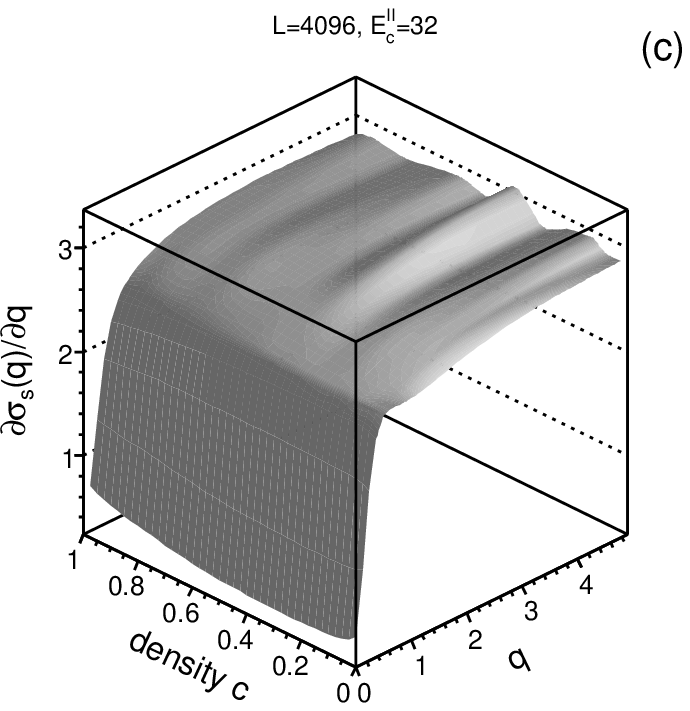}

\caption{\label{fig:dsigma}Plots $\partial\sigma_{s}(q)/\partial q$ versus
exponent $q$ for the lattice size $4096\times4096$, concentrations
$0.0<c\leq1.0$ and thresholds: $E_{c}^{I}=4,$ (a) $E_{c}^{II}=8$,
(b) $E_{c}^{II}=16$, and (c) $E_{c}^{II}=32$.}

\end{figure*}

\begin{figure*}
\includegraphics[width=6cm]{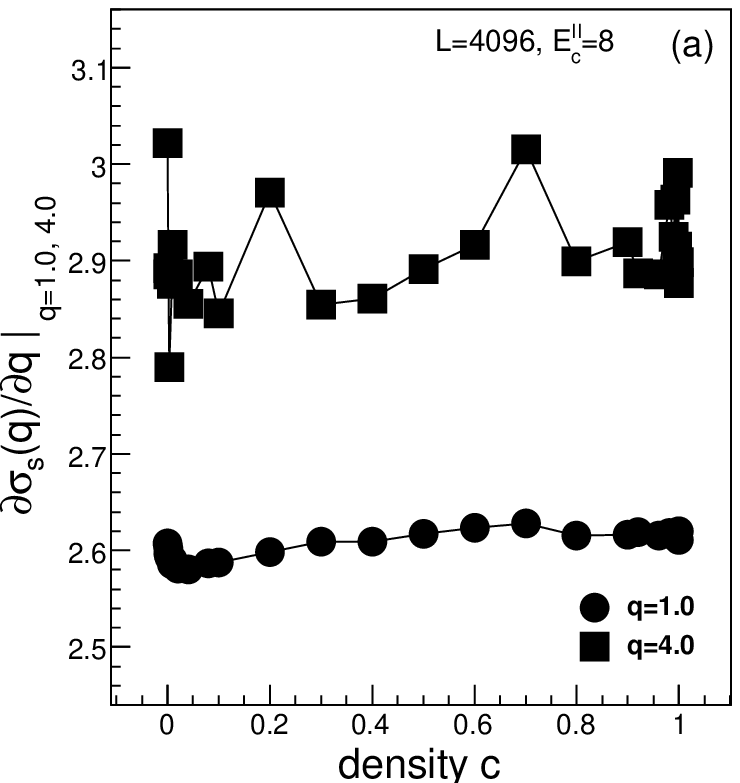}\includegraphics[width=6cm]{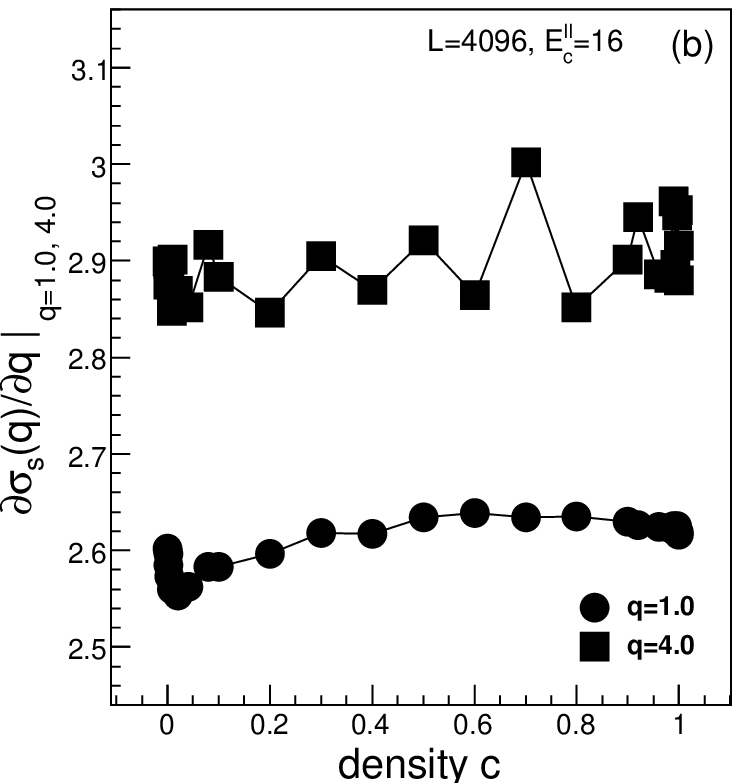}\includegraphics[width=6cm]{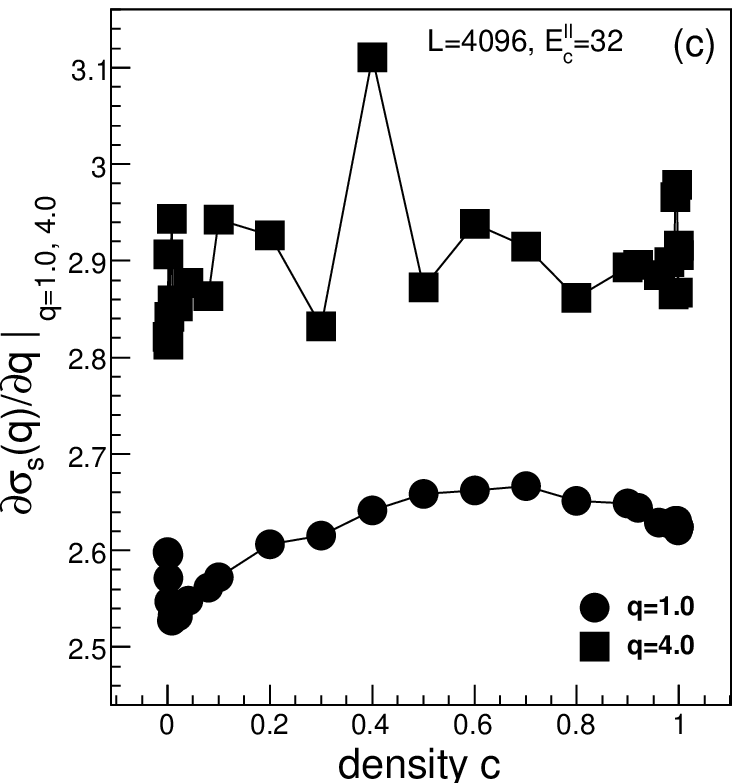}

\caption{\label{fig:dq}The plots $\partial\sigma_{s}(q)/\partial q\mid_{q=1}$
(errors $\pm0.01$) and $\partial\sigma_{s}(q)/\partial q\mid_{q=4}$(errors
$\pm0.03$) versus density $c$ for the linear lattice size $L=4096$
and thresholds $E_{c}^{I}=4$, (a) $E_{c}^{II}=8$, (b) $E_{c}^{II}=16$,
and (c) $E_{c}^{II}=32$.}

\end{figure*}

Using the functions $\sigma_{s}(q)$ I determined the plots $\partial\sigma_{s}(q)/\partial q$
versus the exponent $q$ which are shown in Fig. \ref{fig:dsigma}
for densities $0.0<c\leq1.0$ and thresholds $4<E_{c}^{II}\leq32$.
I have observed that the functions $\partial\sigma_{s}(q)/\partial q$
are increasing if exponents $q$ increase (Fig. \ref{fig:dsigma})
for exponents $q>1.0$, densities $0.0<c<1.0$ and thresholds $8\leq E_{c}^{II}\leq32$.
Surface cuts $\partial\sigma_{s}(q)/\partial q\:\mid_{q=1.0}$ and
$\partial\sigma_{s}(q)/\partial q\:\mid_{q=4.0}$, for exponents $q=1.0$
and $q=4.0$, as functions of density $c$ are shown in Fig. \ref{fig:dq}
to demonstrate this increasing tendency. I have found that $\partial\sigma_{s}(q)/\partial q\:\mid_{q=4.0}>\partial\sigma_{s}(q)/\partial q\:\mid_{q=1.0}$,
$\partial\sigma_{s}(q)/\partial q\:\mid_{q=1.0}\neq const.$ and $\partial\sigma_{s}(q)/\partial q\:\mid_{q=4.0}\neq const.$
(Fig. \ref{fig:dq}) for densities $0.0<c<1.0$ and thresholds $8\leq E_{c}^{II}\leq32$.
This implies that functions $\partial\sigma_{s}(q)/\partial q$ differ
from the function $\partial\sigma_{s}(q)/\partial q$ of the BTW model.
However, for the specific parameters $c=0.0$, $c=1.0$ and $8\leq E_{c}^{II}\leq32$
relaxation rules are precisely balances and the functions $\partial\sigma_{s}(q)/\partial q$
versus $q$ ($q\geq1.0$) are identical within experimental errors
with the function $\partial\sigma_{s}(q)/\partial q$ of the BTW model
\cite{Menech,Karmakar_prl}.

\section{\label{sec:Discussion}Discussion}

I have found that autocorrelations $C(t,\: L,\: c)$ (Fig. \ref{fig:Autocorrelation})
are more complex functions than an autocorrelation of the BTW model
\cite{Menech,Karmakar_prl}. The autocorrelations exhibit oscillating
components (Fig. \ref{fig:Autocorrelation}) which periods and amplitudes
depend on both densities $0.0<c<1.0$ and thresholds $8\leq E_{c}^{II}\leq32$.
The oscillating components are probably caused by a periodicity in
an avalanche wave sequence. I assume that the periodicity could be
a consequence of an excessive energy storing and release in sites
with the thresholds $E_{c}^{II}$, apparently when the sites have
low concentration $c<0.5$. In such conditions, relaxations of these
sites trigger relaxations of surrounding sites, i.e. all surrounding
sites within a certain distance from a disturbing site have to relax
\cite{cer_2002}. This hypothesis could be supported by finding that
for the same time $t$ amplitudes of oscillating components are decreasing
if densities $c$ increase and oscillations disappear near the density
$c=1.0$ (Fig. \ref{fig:Autocorrelation}). The periods were longer
for low densities $c<0.4$ and thresholds $E_{c}^{II}\geq16$ than
for the threshold $E_{c}^{II}=8$. This could be connected with the
stronger effect of disturbing sites $E_{c}^{II}\geq16$ which can
store and release much more energy than disturbing sites $E_{c}^{II}=8$.
However, a nontrivial dependence of periods on thresholds $E_{c}^{II}\geq8$
and densities $c$ (Fig. \ref{fig:Autocorrelation}) and the cause
of period splitting for thresholds $E_{c}^{II}\geq16$ and critical
densities $c$ (Fig. \ref{fig:Autocorrelation} (b) and (c)) are not
understood. For specific densities $c=0.0$ and $c=1.0$, the autocorrelations
$C(t,\: L,\: c=0.0)$ and $C(t,\: L,\: c=1.0)$ agree well with autocorrelation
of the BTW model \cite{Karmakar_prl,Menech}. I have found correlated
avalanche waves under more general conditions in which the relaxation
rules are unbalanced \cite{Karmakar_prl} for densities $0.0<c<1.0$
and thresholds $8\leq E_{c}^{II}\leq32$. This is completely opposite
to the result the hypothesis predicted \cite{Karmakar_prl,Karmakar_pre}.
I think that the hypothesis about a precise relaxation balance \cite{Karmakar_prl,Karmakar_pre}
is valid only for the specific densities $c=0.0$ and $c=1.0$ and
thresholds $8\leq E_{c}^{II}\leq32$.

Two scaling intervals of fluctuations $F(t,\: L=4096)$ (Fig. \ref{fig:Hurst})
support correlated avalanche waves \cite{Menech} for densities $0.0\leq c\leq1.0$
and thresholds $8<E_{c}^{II}\leq32$. The fluctuations $F(t,\: L=4096)$
agree well with a fluctuation of the BTW model only for the densities
$c=0.0$ and $c=1.0$ \cite{Menech,Karmakar_prl} when relaxation
rules are precisely balanced \cite{Karmakar_prl}. For all other
parameters, densities $0.0<c<1.0$ and thresholds $8\leq E_{c}^{II}\leq32$,
relaxation rules are unbalanced and the hypothesis \cite{Karmakar_prl,Karmakar_pre}
predicts a single scaling region with the Hurst exponent $H=1/2$.
However, I have not found single scaling regions of $F(t,\: L=4096)$.
So the fluctuations $F(t,\: L=4096)$ contradict the hypothesis of
precise relaxation balance \cite{Karmakar_prl,Karmakar_pre}. Asymmetries
of functions $H_{1,2}(c)$ to permutations of densities $c$ (Fig.
\ref{fig:Hurst}) could be a consequence of different local effects
of disturbing sites $E_{c}^{II}$ near density $c=0.0$ and disturbing
sites $E_{c}^{I}$ near density $c=1.0$. The Hurst exponents $H_{1}(c)$
are limited by the interval $0.68<H_{1}(c)<0.8$ near the density
$c=0.0$ thus avalanche waves are less correlated than in the BTW
model ($H_{1}=0.8$) and they are more correlated than in the M model
($H=0.5$). The Hursts exponents $H_{2}(c)$ are limited by the interval
$0.44<H_{2}(c)\leq0.56$ which indicate that local perturbation effects
can change a long-term persistence (antipersistence) \cite{Mandelbrot}.
The changes of Hurst exponents $H_{1,2}(c)$ with density $c$ and
threshold $E_{c}^{II}$ demonstrate that local perturbation effects
could change a global wave dynamics. 

I assume that avalanche wave dynamics (autocorrelations and fluctuations)
on a finite size lattices $L\times L$, for $L>L_{c}$ where $L_{c}$
is a critical length, can provide basic information about correlated
(the BTW model) or uncorrelated (the M model) nature of waves in avalanches
not only for the finite size $L$ but also for the size $L$ which
goes to infinity. I assume that $L=4096$ is greater than $L_{c}$.
Then conclusions regarding an avalanche wave dynamics for the finite
lattice size $L=4096$ could be extended for infinite systems. 

A comparison of avalanche structures, for densities $0.0<c<1.0$ and
thresholds $8\leq E_{c}^{II}\leq32$, shows that the avalanche structures
of the model (Sec. \ref{sec:model}) are more disordered than shell-like
structures of the BTW \cite{Hur,Karmakar_prl} model but they are
more ordered than structures of the M model \cite{Hur,Karmakar_prl}.
Only for the specific densities $c=0.0$ and $c=1.0$, the avalanche
structures are exactly shell-like as well as in the BTW model \cite{Hur,Karmakar_prl}.
I assume that connection between avalanche structures and autocorrelations
near the density $c=0.5$ are more weak for the model (Sec. \ref{sec:model})
than for the BTW or M models \cite{Hur,Karmakar_prl}. Because avalanche
structures are more disordered (Fig. \ref{fig:avalanche}) than structures
of the BTW model and more ordered than structures of the M model,
however avalanche waves are correlated as in Fig. \ref{fig:Autocorrelation}. 

A hypothesis about a precise relaxation balance \cite{Karmakar_prl}
for unbalanced relaxation rules predicts FSS where $\partial\sigma_{s}(q)/\partial q=const.$
for all exponents $q>1.0$. In such case, an expectation $\partial\sigma_{s}(q)/\partial q\:\mid_{q=4}\doteq\partial\sigma_{s}(q)/\partial q\:\mid_{q=1}$
is correct. However, avalanche size moments do not support this expectation,
because I have found that $\partial\sigma_{s}(q)/\partial q\:\mid_{q=4}>\partial\sigma_{s}(q)/\partial q\:\mid_{q=1}$
and functions $\partial\sigma_{s}(q)/\partial q\:\mid_{q=4}$ and
$\partial\sigma_{s}(q)/\partial q\:\mid_{q=1}$ (Fig. \ref{fig:dq})
depend on a density $c$ for all densities $0.0<c<1.0$ and thresholds
$8\leq E_{c}^{II}\leq32$. This is a reason why functions $\partial\sigma_{s}(q)/\partial q$
of the model (Sec. \ref{sec:model}) cannot be identical with the
function $\partial\sigma_{s}(q)/\partial q$ of the BTW model \cite{Karmakar_prl,Manna}
($c=0.0$ and $1.0$) for densities $0.0<c<1.0$. Although, I conclude
that avalanche waves are correlated in avalanches and avalanche sizes
scale as a multifractal for all investigated parameters.

For specific densities $c=0.0$ and $1.0$, the basic assumptions
$H_{i}-H_{i}^{'}=0$ of the precise relaxation balance hypothesis
\cite{Karmakar_prl} are correct. However, all results (Sec. \ref{sec:Results})
show a multifractal scaling of avalanche sizes not only for the specific
conditions but also for the more general conditions $0.0<c<1.0$ and
$8\leq E_{c}^{II}\leq32$ in which $|H_{i}-H_{i}^{'}|=4n$, $0\leq n\leq7$
where $n$ is a natural number. Thus I conclude that the hypothesis
about precise relaxation balance \cite{Karmakar_prl} is valid only
for the specific parameters $c=0.0$ and $1.0$ and $E_{c}^{II}>4$.

I have not observed a transition from multifractal scaling to FSS
of avalanche sizes which has been expected for low densities near
$c=0.0$ (Sec. \ref{sec:Introduction}). However, the autocorrelations,
Hurst exponents and avalanche size moments support the previous conclusions
\cite{cer_2002,cer_2008,Cernak_2006} that multifractal scaling is
very sensitive to local perturbations for densities near $c=0.0$. 

Does the model belong to the BTW class? The answer is not uniform.
The model shows common features with the BTW model for example, correlated
waves in avalanches and multifractal scaling of avalanche sizes. However,
I have demonstrated that the autocorrelations (Fig. \ref{fig:Autocorrelation})
are complex functions, Hurst exponents are functions of densities
$c$ (Fig. \ref{fig:Hurst}) and holes can be found in avalanches
while all these features are not typical for the BTW model. A possible
difference between the inhomogeneous sand pile model \cite{cer_2002}
and the BTW model \cite{BTW} is supported by functions $\partial\sigma_{s}(q)/\partial q$
(Figs. \ref{fig:dsigma} and \ref{fig:dq}) which are not identical
with the function $\partial\sigma_{s}(q)/\partial q$ of the BTW model
except the specific densities $c=0.0$ and $1.0$.

\section{\label{sec:Conclusion}Conclusion}

Applying the classification scheme \cite{Karmakar_pre} on the inhomogeneous
sandpile model model (Sec. \ref{sec:model}) places the model in the
M universality class. However, I have demonstrated (Sec. \ref{sec:Results})
that the model belongs neither in the M universality class nor in
the BTW universality class (Sec. \ref{sec:Discussion}). I assume
that it belongs in a multifractal universality class \cite{Tebaldi}
which is more general than the BTW class \cite{cer_2008}. Based
on the wave autocorrelations, fluctuations, avalanche structures and
avalanche size moments, I conclude that an avalanche wave dynamics
and avalanche size scaling depend on local relaxation details. In
addition, a hypothesis about precise relaxation balance \cite{Karmakar_prl,Karmakar_pre}
could be a specific case of a more general rule. The reason why a
multifractal scaling is very sensitive to local perturbations is not
well understood. 

\

\begin{acknowledgments}
The author thanks A. Read for his comments to the manuscript. Computer
simulations were carried out in the projects NorduGrid and KnowARC.
This work was supported by Slovak Research and Development Agency
under contact No. RP EU-0006-06. 
\end{acknowledgments}

\end{document}